\documentclass[prb,twocolumn,superscriptaddress]{revtex4-1}
\usepackage{graphicx}
\usepackage{dcolumn}
\usepackage{bm}
\usepackage{color}
\usepackage{amsmath, amsthm, amssymb,stmaryrd}
\usepackage[retainorgcmds]{IEEEtrantools}
\usepackage{polski}
\usepackage[polish,french,english]{babel}
\newcommand{\rto}[1]{$R_2$Ti$_2$O$_7${#1}}
\newcommand{\tto}[1]{Tb$_2$Ti$_2$O$_7${#1}}
\newcommand{\hto}[1]{Ho$_2$Ti$_2$O$_7${#1}}
\newcommand{\dto}[1]{Dy$_2$Ti$_2$O$_7${#1}}

\usepackage{amssymb}
\usepackage{epstopdf}
\newcommand{\ra}[1]{\renewcommand{\arraystretch}{#1}}

\begin{document}

\title{Crystal field parameters of the rare earth pyrochlores $R_2$Ti$_2$O$_7$ ($R$ = Tb, Dy, Ho)}
\author{M Ruminy}
\email{martin.ruminy@gmx.de}
\affiliation{Laboratory for Neutron Scattering and Imaging, Paul Scherrer Institut, 5232 Villigen PSI, Switzerland}
\author{E Pomjakushina}
\affiliation{Laboratory for Scientific Developments \& Novel Materials, Paul Scherrer Institut, 5232 Villigen PSI, Switzerland}
\author{K Iida}
\affiliation{Comprehensive Research Organization for Science and Society (CROSS), Tokai, Ibaraki 319-1106, Japan}
\author{K Kamazawa}
\affiliation{Comprehensive Research Organization for Science and Society (CROSS), Tokai, Ibaraki 319-1106, Japan}
\author{D T Adroja}
\affiliation{ISIS Facility, Rutherford Appleton Laboratory, Chilton, Didcot, Oxon OX11 0QX, United Kingdom}
\affiliation{Highly Correlated Electron Group, Physics Department, University of Johannesburg, P.O. Box 524, Auckland Park 2006, South Africa}
\author{U Stuhr}
\affiliation{Laboratory for Neutron Scattering and Imaging, Paul Scherrer Institut, 5232 Villigen PSI, Switzerland}
\author{T Fennell}
\email{tom.fennell@psi.ch}
\affiliation{Laboratory for Neutron Scattering and Imaging, Paul Scherrer Institut, 5232 Villigen PSI, Switzerland}

\date{\today}

\begin{abstract}
In this work we present inelastic neutron scattering experiments which probe  the single ion ground states of the rare earth pyrochlores $R_2$Ti$_2$O$_7$ ($R$ = Tb, Dy, Ho).  \dto{} and \hto{} are dipolar spin ices, now often described as hosts of emergent magnetic monopole excitations; the low temperature state of \tto{} has features of both spin liquids and spin glasses, and strong magnetoelastic coupling.  We measured the crystal field excitations of all three compounds and obtained a unified set of crystal field parameters.  Additional measurements of a single crystal of \tto{} clarified the assignment of the crystal field levels in this material and also revealed a new example of a bound state between a crystal field level and an optical phonon mode.
\end{abstract}

\pacs{}
\maketitle

\section{\label{sec:Introduction}Introduction}

The counterintuitive demonstration that a system of Ising-like spins interacting ferromagnetically on the pyrochlore lattice is frustrated (in the case that the local Ising axes are the $\langle 111\rangle$ directions), while the antiferromagnetic counterpart is unfrustrated~\cite{Harris:1997uy,Bramwell:1998wk,Moessner:1998uj}, underpins the ongoing interest in the three rare earth pyrochlores \tto{}, \dto{}, and \hto{}~\cite{Gardner:2010fu}.  The latter two are model materials for the physics of dipolar spin ice - a microscopic model derived from the $\langle111\rangle$-Ising ferromagnet (or near-neighbor spin ice) by the incorporation of dipolar interactions~\cite{denHertog:2000tc,Bramwell:2001tpa}, and now developed into an effective theory of a Coulomb gas of emergent magnetic monopoles~\cite{Castelnovo:2008hb,Ryzhkin:2005ko,Castelnovo:2012kk}.  The monopole charge is revealed by the construction of the so-called dumbbell model~\cite{Castelnovo:2008hb} where each magnetic moment is replaced by a dumbbell carrying a magnetic charge $\pm q$ at each end.  The magnitude of the charge can be derived from the size of the magnetic moment $\mu$, and the separation of the tetrahedron centers.  Current activity is focussed on developing understanding of magnetic Coulomb gases, and testing the applicability of these ideas in real materials~\cite{Jaubert:2009ed,Zhou:2011fq,Giblin:2011ft,Castelnovo:2011dp,Revell:2012hq,Bramwell:2012kw,Kaiser:2013di,Kaiser:2015ty}.  \tto{} on the other hand is ostensibly an example of the unfrustrated antiferromagnetic case, but strongly confounds the expected scenario of simple and complete long-range magnetic order by remaining in a correlated but magnetically disordered state down to the lowest temperatures.  Experimental~\cite{Baker:2012cd,Legl:2012gm,Lhotel:2012hb,Fennell:2012ci,Petit:2012ko,Fritsch:2013dv,Taniguchi:2013fi,Sazonov:2013cp,Yin:2013iz,Fennell:2014gf,Fritsch:2014cx,Guitteny:2013hf,Guitteny:arx} and theoretical~\cite{Kao:2003ex,Enjalran:2004kr,Molavian:2007ve,Molavian:2009dl,Bonville:2011dw,Curnoe:2007hd,Curnoe:2008gy,Klekovkina:2011vv,Curnoe:2013iz,Sazonov:2013cp,Bonville:2014eq,Jaubert:2015fm,Kadowaki:2015_1,Kadowaki:2015_2} attempts to cast light on the mechanism for this exception are the main activities. 

In magnetic materials based on rare earth ions, the crystal electric field (CEF) controls the single ion ground state, which determines the size and anisotropy of the magnetic moment; the wavefunctions of the single ion ground and excited states determine transverse components important for quantum fluctuations~\cite{Gingras:2014ip,Rau:2015jj}, spin tunneling~\cite{Tomasello:2015vg,Rau:2015jj}, modification of spin interactions or anisotropies by virtual fluctuations~\cite{Molavian:2007ve,McClarty:2009gn,Petit:2014jb}, and the presence of potentially interesting higher multipole moments~\cite{Petit:2012ko,Huang:2014kq,Kadowaki:2015_1,Kadowaki:2015_2}; the spectrum of excited CEF levels determines the temperature dependence of all these quantities, and controls interactions with phonons for spin flipping mechanisms~\cite{Finn:2002fs}.  Hence, in a spin ice such as \dto{} or \hto{} it is essential to understand the CEF Hamiltonian in order to quantify the contribution of the magnetic moment $\mu$ to the monopole charge, to understand mechanisms by which the monopoles can hop~\cite{Tomasello:2015vg}, and to quantify possible quantum corrections to the classical model~\cite{McClarty:2014vc,Rau:2015jj}.  In \tto{}, understanding of the CEF wavefunctions is essential for development of theories of virtual fluctuations~\cite{Molavian:2007ve}, magnetoelastic interactions~\cite{MAMSUROVA:1988wg,Klekovkina:hc,Fennell:2014gf,Bonville:2014eq}, or higher multipoles~\cite{Kadowaki:2015_1,Kadowaki:2015_2,Guitteny:2013hf}, which are all viewed as possible means to suppress long range magnetic order.  

Because the CEF parameters of a rare earth-based magnet are so important, various determinations have been made before for all three compounds. In \hto{} and \dto{} it is known from magnetization measurements that the single ion ground states must be close to pure $|m_J=\pm8\rangle$ and $|m_J=\pm15/2\rangle$ doublets respectively~\cite{Bramwell:2000tc}, and for \hto{} this was confirmed by inelastic neutron scattering~\cite{Rosenkranz:2000te}.  The strong anisotropy, and also the activation energy for thermal spin flips ($\approx 293$ K for \hto{}~\cite{Ehlers:2002jz}, and in the range $200-300$ K for \dto{}~\cite{Snyder:2003ek}) points to a very large gap to the first excited crystal field state.  However, the exact level scheme and wavefunctions of \dto{} are still not known.  In the absence of neutron scattering results, CEF parameters for \dto{} have been obtained by point-charge scaling of the known CEF parameters of \hto{}~\cite{Qiu:2002}, or from a CEF analysis with simultaneous point-charge scaling of the entire $R_2$Ti$_2$O$_7$ series~\cite{Malkin:2004ke,Bertin:2012gu}, procedures which provide, at best, a useful guide to the actual values.  

In \tto{} there is considerable activity associated with the determination of the CEF parameters.  Several reports concluded that the ground state doublet was separated from the first excited state by only 15 K, and that these were dominated by $|m_J=\pm4\rangle$ and $|m_J=\pm5\rangle$ components respectively~\cite{aleksandrov,Gingras:2000vd,Gardner:2001kv,Mirebeau:2007hi}.  However, the double-peak shape of a magnetic excitation at 15.5 meV was not previously noticed, and its assignment in more recent studies has led to debate.  Because of the peak shape, the single level was first reinterpreted as two crystal field excitations~\cite{Zhang:2014ef}.  The resulting CEF parameters reversed the dominant contributions to the wavefunctions of the ground and first excited states.  However, it was pointed out~\cite{Princep:2015kt} that the new parameters  were inconsistent with all other proposals and the entire feature was subsequently treated as a single crystal field level with a splitting.  The reason for the splitting could not be established experimentally, though it has been proposed theoretically to be due to a type of coupled electron-phonon state~\cite{Klekovkina:hc}.  

By measuring the spectra of all three compounds, taking advantage of our investigation of the phonon band structure of the rare earth titanates~\cite{phonons}, and by investigating the double peak feature in \tto{} using a single crystal, we are able to clarify the assignments and propose a set of CEF parameters which are consistent across the three compounds.  This also constitutes the first direct determination of the energy levels and CEF parameters for \dto{}.  Furthermore, we identify the double peak in \tto{} as the signature of a crystal field-phonon bound state (sometimes known as a vibron bound state), somewhat similar to that in CeAl$_2$~\cite{Thalmeier:1982wc} or CeCuAl$_3$~\cite{Adroja:2012cm}.

\section{\label{sec:Experimental_methods}Experimental methods}

The samples and spectrometers used in this study have already been described elsewhere~\cite{phonons}.  We repeat the details for completeness.

Inelastic neutron time-of-flight measurements on 10 g powder samples of \hto{} and \tto{} were performed on the MERLIN spectrometer at ISIS \cite{Bewley:2006}.  The samples were packed in envelopes of aluminum foil which were curled up to form an annular cylinder with diameter and height of 40\,mm. Subsequently, the samples were sealed into aluminum cans containing helium exchange gas, and cooled by a closed-cycle refrigerator on the instrument.  Different settings with incoming neutron energies of $E_i=30$ (only for $R=$Tb), $60$ and $150$\,meV, and corresponding chopper frequencies of $f=200$, $400$, and $600$\,Hz were chosen to record data at $T=5, 50,$ and $200$\,K for $400$\,{$\mu$}Amp hrs ($\approx 2.5$ hours at ISIS full power) each. The instrumental background at MERLIN is very low, and no aluminum contribution was visible in the raw data, so we did not measure the empty sample can separately. The raw data were corrected for detector-efficiency using a vanadium reference sample. 

A powder sample of \dto{} (with natural dysprosium isotopic abundance) was investigated using the 4SEASONS spectrometer at J-PARC~\cite{Kajimoto:2011ja}.  The 5 g sample was packed in an aluminum foil envelope which was wrapped into a cylinder of 30 mm diameter and 50 mm height, and then sealed in an aluminum can with helium exchange gas.  The thickness of the sample was carefully controlled so as not to exceed 0.5 mm, to maximize the inelastic signal despite the large absorption cross section of natural dysprosium.  Using tabulated values for the total scattering and absorption cross sections of the constituent elements of \dto{}, we estimated an optimal thickness of 0.85 mm for $E_i=17$ meV, making the common assumption that the packed powder has a density approximately 0.6 of the crystallographic density.  When rolled into a cylinder, the sample thickness traversed by the neutrons approximately doubles, and our envelope thickness of $<0.5$ mm gave a total path in the sample close to the optimum value.  Magnetic scattering was not included, but can make only a small contribution given the weight of the very large coherent cross section of dysprosium in the calculation (due to $^{164}$Dy).  We note that the absorption cross section is very significantly reduced at the higher energies in which we were mostly interested, for example $\sigma_{\mathrm{abs}}^{E=150 {\mathrm{meV}}}=0.4\sigma_{\mathrm{abs}}^{E=25 {\mathrm{meV}}}$ (absorption cross sections are typically tabulated for $E=25$ meV neutrons).  For energies like 55 or 150 meV our sample was therefore below optimal thickness, resulting in high transmission but longer counting times.

4SEASONS was operated in repetition rate multiplication mode~\cite{Nakamura:2009iw}.  Using a Fermi chopper frequency of 250 Hz, the phases of the other choppers were configured so that for a single source pulse, spectra were recorded either for $E_i=153.2, 55.4, 28.3, 17.1$ meV, or for $E_i=308.7, 82.0, 37.1, 21.1$ meV simultaneously.  Measurements were taken at $T = 5$ K in both settings, and at $T = 200$ K in the $E_i = 153.2$ meV setting, for 8 hours each.  In the $E_i = 153.2$ meV setting, the instrumental background was measured at both temperatures to subtract the significant contribution from scattering due to phonons of the aluminum sample can from the raw data. The raw data were corrected for detector-efficiency using a vanadium reference sample~\cite{Inamura:2013fe}.  Absorption corrections were found to be unnecessary due to the optimized transmission described above.

The assignment of optical phonons and crystal field excitations between 8 and 20 meV in \tto{} was further investigated using the thermal triple-axis neutron spectrometer EIGER at the Swiss neutron spallation source SINQ. The single crystal sample was previously characterized by heat capacity, x-ray diffraction, and inelastic neutron scattering measurements.  This characterization was described in Ref.~[\onlinecite{mem_all_samps}] (in which the sample is known as EP2), where the sample was shown to have no heat capacity peak at $T\approx 0.5$ K, and the composition was estimated to be Tb$_{2.04}$Ti$_{1.96}$O$_{6.98}$.  The crystal, which has a mass of approximately 1.2\,g, was fixed by aluminum wires on an aluminum holder such that the scattering plane was spanned by $(h,h,l)$ wavevectors, then mounted in a standard helium cryostat.  The spectrometer was operated with fixed final neutron energy $E_f = 14.7$ meV using the $(0,0,2)$ reflection of the pyrolitic graphite (PG) monochromator and analyzer.  A PG filter in the scattered beam was used to eliminate contamination by scattering of neutrons with higher order wavelengths. Constant wavevector scans ($E$-scans) were performed in the energy transfer window $E = [8, 20]$ meV at Brillioun zone (BZ) center and boundary points along the three high symmetry directions to measure the $\vec{Q}$ and temperature dependence of the components of the broad envelope centered at $E\approx 15.5$ meV.  Using the $(0,0,4)$ reflection of the monochromator, $E$-scans with improved energy resolution (but 70 \% reduced signal) were performed at selected $\vec{Q}$-points.  

\section{\label{sec:cef_results}Crystal field spectra: Results}

\begin{figure*}
\includegraphics[scale=1]{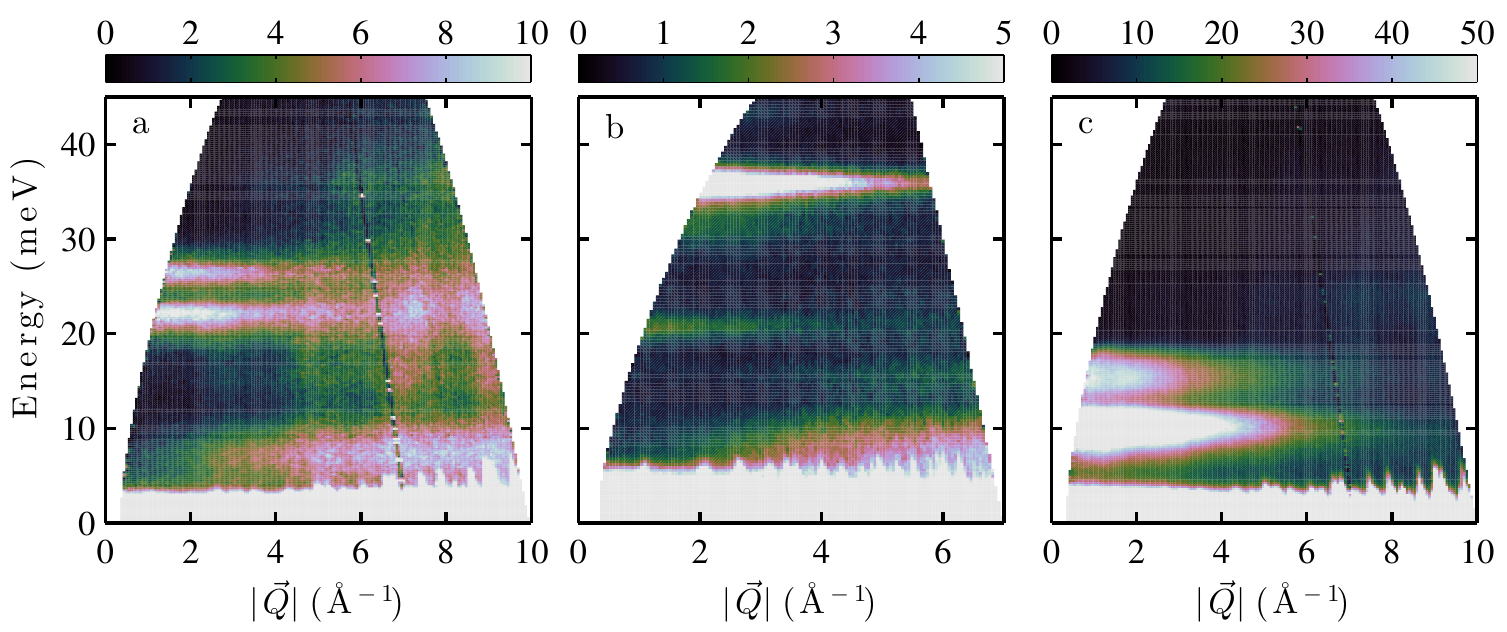}
\caption{Inelastic neutron spectra of polycrystalline samples of \hto{} (a), \dto~(b) and \tto~(c) at $T=5$\,K. The neutron intensities are represented in colormaps of arbitrary scale. Dispersionless excitations with highest intensity at lowest momentum transfers $|\vec{Q}|$ are magnetic excitations from the CEF ground state to excited CEF states. Intensities at large $|\vec{Q}|$ are due to scattering by phonons. The first two CEF excitations in \tto{} are particularly intense, and hence appear broad due to the cutoff of the intensity scale.  Differences in the angular coverage of MERLIN (a,c) and 4SEASONS (b) are responsible for the different shapes of the projection of the data in the $(|\vec{Q}|,E)$-plane.}
\label{fig:cef_maps}
\end{figure*}

Fig.~\ref{fig:cef_maps} summarizes the neutron spectra of all three rare earth titanate pyrochlores recorded with intermediate incident energy ($E_i\approx60$ meV), at $T=5$ K.  The CEF transitions manifest themselves as dispersionless excitations with highest intensity at lowest momentum transfer.  Intense transitions from the CEF ground states of all three titanates are well resolved.

In \hto{}, in the $D_{3d}$ point group of the rare earth site in the pyrochlore structure, in the paramagnetic state, the free ion ground state multiplet $^5I_8$ of the Ho$^{3+}$ ion splits into five $E_g$ doublets, three $A_{1g}$ and two $A_{2g}$ singlets, whose energies were first determined by Rosenkranz {\it et al.}~\cite{Rosenkranz:2000te}. Panel a of Fig.~\ref{fig:cef_maps} shows the first two ground state transitions at $E\approx21.9$ meV and $E\approx26.3$ meV.  A third, weak, CEF transition is found in the tail of the latter excitation. In total we observed six ground state transitions (the others are not covered by the colormap), as well as various levels which can be located by the energies of transitions amongst excited states once the lowest states are thermally populated.  The  transitions are clearly magnetic as their intensity follows the dipole form factor of the Ho$^{3+}$ ion (not shown).  The observed spectrum is completely consistent with the observations of Rosenkranz {\it et al.}~\cite{Rosenkranz:2000te}.  The energies and intensities of the transitions are tabulated in Table~\ref{tab:cf_fitting}.

\begin{figure}
\includegraphics[scale=1]{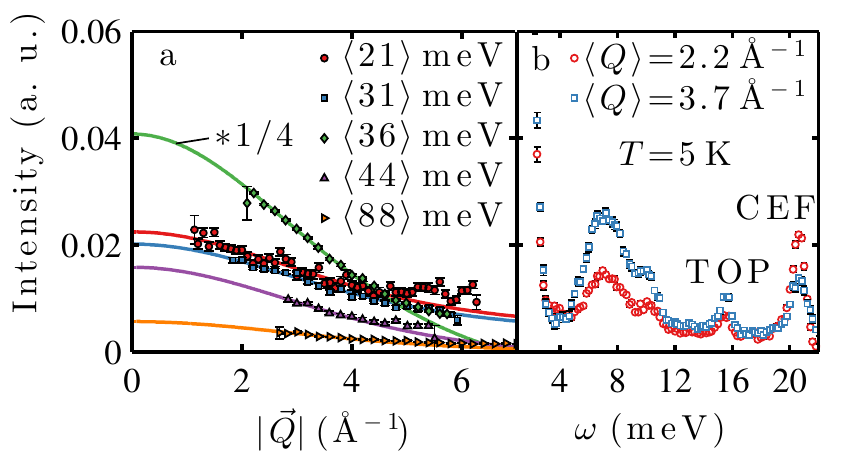}
\caption{Details of the measured CEF spectrum in \dto{}. Panel a: $|\vec{Q}|$ dependence of the CEF excitations at $T=5$\,K. The data points represent an integrated energy band centered at the indicated energy transfer with band widths of $\pm1$ (9)\,meV for $E_i=55$ (150)\,meV (two of the six observed transitions are incorporated in the integration at 88 meV). The lines are fits to the dipolar form factor of Dy$^{3+}$ ions, confirming the magnetic nature of the excitations. Panel b: Integration of raw data measured with $E_i=28$\,meV at low and large momentum transfers ($\Delta|\vec{Q}|=0.4$ \AA$^{-1}$) to distinguish scattering of magnetic and phononic origins. Inelastic scattering at low energy transfers stems dominantly from acoustic phonons, and the sharp feature at $E\approx15.5$\,meV is due to a nearly dispersionless transverse optic phonon (TOP)~\cite{phonons}. The first CEF excitation appears at $E\approx21$\,meV.}
\label{fig:dto_details}
\end{figure}

Because Dy$^{3+}$ is a Kramers ion, its free ion ground state term $^6H_{15/2}$ splits into 8 $E_u$ doublets in the paramagnetic state.  Our measurement with $E_i=300$ meV excludes CEF transitions at energy transfers larger than 100 meV.  We clearly observed six of the expected seven transitions below 100 meV, with the energies summarized in Table~\ref{tab:cf_fitting}.  As shown in Fig.~\ref{fig:cef_maps}b, the first excited CEF doublet in \dto{} appears at an energy transfer of $E\approx20.9$ meV.  The $|\vec{Q}|$-dependence of these modes agrees exactly with the dipole form factor of the Dy$^{3+}$ ion, as illustrated in Fig.~\ref{fig:dto_details}a.  The seventh transition is predicted to be very weak, and to lie close to the two highest observed transitions, from which we cannot resolve it.  We also observe another weak dispersionless feature at $E\approx15$ meV whose width is similar to that of a CEF excitation, but whose intensity increases with momentum transfer (see Fig.~\ref{fig:dto_details}b).  The latter is an obvious signature of scattering from phonons, and our phonon calculations~\cite{phonons} confirm the presence of a dispersionless transverse optic phonon (TOP) at this energy transfer in the $R_2$Ti$_2$O$_7$ phonon spectrum. As we will see below, this mode has important consequences for \tto{}, but in \dto{} it is isolated from the CEF states and can be clearly observed as a pure phonon.

In \tto{}, the free ion ground state multiplet $^7F_6$ of the Tb$^{3+}$ ion splits into four $E_g$ doublets and five singlets (three $A_{1g}$ and two $A_{2g}$) in the paramagnetic state.  Our neutron time-of-flight data show identical features to those seen and analyzed in Ref.~[\onlinecite{Princep:2015kt}].   Strong CEF excitations appear at $E\approx10.2$ meV and $E\approx49$ meV.  At $E\approx61$ meV, a very weak CEF excitation was observed.  An intense double-peak shaped magnetic excitation appears at $E\approx15.5$ meV, the assignment of which has recently been debated~\cite{Zhang:2014ef,Princep:2015kt}.  We have examined it further by experimenting on single crystals and present that investigation below.  An important conclusion from the single crystal experiment is that there is only one CEF excitation in the double peak feature, and it is the second peak at $E\approx16.7$ meV.  All the observed CEF excitations follow the dipole form factor of Tb$^{3+}$ (not shown).  From our extensive single crystal measurements on \tto{} we also know of the very strong first excited CEF doublet centered at $E\approx1.5$ meV, which is not resolved in the present neutron time-of-flight measurements.  We will subsequently refer to the first three excitations of \tto{} as CEF1 ($E\approx1.5$ meV), CEF2 ($E\approx10.2$ meV), and CEF3 ($E\approx16.7$ meV).

\begin{table}
  \caption{ \label{tab:cf_fitting} Comparison of observed and calculated energies and neutron intensities of CEF excitations with given symmetry in \rto{} at 5 K. For each of the three rare earth titanates, the intensities are presented relative to the intensity of a well-resolved, intense, CEF excitation.}
  \centering
  \ra{1.3}
  \begin{tabular}{c c c c c c c c}\hline \hline 
      \multicolumn{8}{l}{\hto{}} \\ \hline
  \multicolumn{2}{c}{Symmetry} & $E_{\rm obs}$ & $E_{\rm LS}$ & $E_{\rm IM}$ & $I_{\rm obs}$ & $I_{\rm LS}$ & $I_{\rm IM}$ \\ \hline
  \multicolumn{2}{c}{$E_g$   } &  0.0          &  0.0         &  0.0         & -             & 43.5         & 47.9  \\
  \multicolumn{2}{c}{$A_{2g}$} & -             & 20.1         & 20.8         & -             & 0.16         & 0.02  \\
  \multicolumn{2}{c}{$E_g$   } & 21.9(2)       & 21.9         & 21.9         & 1.00          & 1.00         & 1.00  \\
  \multicolumn{2}{c}{$E_g$   } & 26.3(2)       & 26.3         & 26.3         & 0.7(1)        & 0.94         & 0.64  \\
  \multicolumn{2}{c}{$A_{1g}$} & 28.3(4)       & 28.2         & 27.9         & 0.1(1)        & 0.20         & 0.26  \\
  \multicolumn{2}{c}{$E_g$   } & 61.0(3)       & 60.6         & 61.0         & 5(1)          & 4.61         & 5.39  \\
  \multicolumn{2}{c}{$A_{1g}$} & -             & 70.9         & 71.6         & -             & 0.12         & 0.29  \\
  \multicolumn{2}{c}{$A_{2g}$} & -             & 71.1         & 72.4         & -             & 0.20         & 0.14  \\
  \multicolumn{2}{c}{$E_{g}$ } & 72(1)         & 73.2         & 73.0         & 0.4(2)        & 0.03         & 0.07  \\
  \multicolumn{2}{c}{$E_g$   } & 78.7(4)       & 78.8         & 78.9         & 0.5(1)        & 0.57         & 0.51  \\
  \multicolumn{2}{c}{$A_{1g}$} & -             & 82.7         & 82.2         & -             & 0.01         & 0.04  \\ \hline\hline
    \multicolumn{8}{l}{\dto{}} \\ \hline
  \multicolumn{2}{c}{Symmetry} & $E_{\rm obs}$ & $E_{\rm LS}$ & $E_{\rm IM}$ & $I_{\rm obs}$ & $I_{\rm LS}$ & $I_{\rm IM}$ \\ \hline
  \multicolumn{2}{c}{$E_u$   } &  0.0          &  0.0         &  0.0         & -             & 8.22         & 8.46  \\
  \multicolumn{2}{c}{$E_u$   } & 20.9(4)       & 21.1         & 21.0         & 0.12(2)       & 0.09         & 0.09  \\
  \multicolumn{2}{c}{$E_u$   } & 30.9(4)       & 31.1         & 30.6         & 0.05(2)       & 0.02         & 0.03  \\
  \multicolumn{2}{c}{$E_u$   } & 36.0(1)       & 36.0         & 36.0         & 1.00          & 1.00         & 1.00  \\
  \multicolumn{2}{c}{$E_u$   } & 43.6(5)       & 43.1         & 43.7         & 0.06(2)       & 0.06         & 0.09  \\
  \multicolumn{2}{c}{$E_u$   } & 83.4(10)      & 85.1         & 83.7         & 0.08(2)       & 0.03         & 0.03  \\
  \multicolumn{2}{c}{$E_u$   } & -             & 88.2         & 87.8         & -             & 0.03         & 0.02  \\
  \multicolumn{2}{c}{$E_u$   } & 92.5(15)      & 90.3         & 90.9         & 0.05(2)       & 0.01         & 0.01 \\ \hline\hline
  \multicolumn{8}{l}{\tto{}} \\ \hline
  Sym. (LS)& Sym. (IM)& $E_{\rm obs}$ & $E_{\rm LS}$ & $E_{\rm IM}$ & $I_{\rm obs}$ & $I_{\rm LS}$ & $I_{\rm IM}$ \\ \hline
  $E_g$    & $E_g$    & 0.0           &  0.0         &  0.0         & -             & 0.80         & 1.42  \\
  $E_g$    & $E_g$    & 1.5(1)        &  1.6         &  1.5         & 1.5(2)        & 2.18         & 1.49  \\
  $A_{2g}$ & $A_{2g}$ & 10.2(2)       & 10.4         & 10.2         & 1.00          & 1.00         & 1.00  \\
  $A_{1g}$ & $A_{1g}$ & 16.7(4)       & 16.2         & 17.0         & 0.3(1)        & 0.27         & 0.51  \\
  $A_{2g}$ & $E_g$    & 42(2)         & 44.1         & 38.7         & 0.01(1)       & 0.02         & 0.02  \\
  $A_{1g}$ & $A_{2g}$ & -             & 45.3         & 47.9         & -             & 0.21         & 0.01  \\
  $E_g$    & $A_{1g}$ & 49(1)         & 45.7         & 48.4         & 0.1(1)        & 0.01         & 0.07  \\
  $E_g$    & $E_g$    & 61(2)         & 62.9         & 60.2         & 0.05(2)       & 0.05         & 0.02  \\
  $A_{1g}$ & $A_{1g}$ & -             & 69.8         & 70.4         & -             & 0.00         & 0.00  \\ \hline\hline
  \end{tabular}
\end{table}

We extracted the energy transfers and relative intensities of the transitions from the $T=5$ K data sets. The $T=200$ K data sets serve as an additional verification of the obtained CEF Hamiltonians, in particular in the context of observed and calculated excitations between thermally excited CEF states. By integrating the inelastic powder spectrum over different ranges of momentum transfers we distinguished scattering intensities with magnetic and phononic origin. The $|\vec{Q}|$-integration of $S(|\vec{Q}|,\omega)$ for dominantly magnetic signal ranged on average from $0<|\vec{Q}|<4$ \AA$^{-1}$, and for signal from lattice vibrations from $|\vec{Q}|>8$ (6) \AA$^{-1}$ for data from MERLIN (4SEASONS). Features appearing in both integrated spectra and unambiguously originating from scattering by $R_2$Ti$_2$O$_7$ phonons were used to estimate a scaling factor between the spectra of small and large momentum transfers. The elastic line in the latter spectrum was eliminated before the scaled phonon contribution was subtracted from the low-$|\vec{Q}|$ spectrum.   Transition energies and integrated intensities were determined by fitting an asymmetric pseudo-Voigt lineshape to the peaks of the phonon-subtracted spectra at $T = 5$ K. The parameters of the asymmetric lineshape were determined from the shapes of the incoherent elastic lines, and applied to all observed magnetic peaks. With this procedure we obtained one unique set of peak shape parameters for each $E_i$-setting of MERLIN, while the description of the lineshape of 4SEASONS requires different parameters, as expected. Our experimental observations of the CEF energies and intensities relative to a well-resolved excitation are summarized in Table \ref{tab:cf_fitting} for all three compounds.

Based on these observations, the six CEF parameters $B_q^k$ (Wybourne normalization) of the rare earth pyrochlore CEF Hamiltonian,
 \begin{IEEEeqnarray}{rCl}
\mathcal{H}_\mathrm{CEF}&=&B_0^2C_0^2 + B_0^4C_0^4 +B_3^4(C_{-3}^4 - C_3^4)+B_0^6C_0^6\nonumber \\
 & & + B_3^6(C_{-3}^6-C_3^6) + B_6^6(C_{-6}^6 + C_6^6),
 \end{IEEEeqnarray}
were refined using the program Spectre~\cite{spectre} using both the $LS$-coupling and intermediate (IM) coupling  schemes.   $LS$-coupling, in which the Coulomb repulsion dominates the spin-orbit interaction (i.e. in the Russell-Saunders approximation) and the states come only from the lowest $^{2S+1}L_J$ multiplet, is usually appropriate for rare earth ions. If the spin-orbit interaction is dominant (which is not the case for rare earth ions), $j-j$ coupling would be more appropriate.  Between these limits where there is no specific hierarchy for the interactions, the intermediate coupling scheme, which considers both interactions and the full $f^n$ configuration can be appropriate.  Intermediate coupling mixes contributions from higher multiplets into crystal field wave functions, and its likely applicability can therefore also be gauged by the ratio of ground multiplet splitting and lowest excited multiplet energies.  The importance of intermediate coupling for \tto{} has recently been justified~\cite{Princep:2015kt}, and we have obtained wavefunctions using both the $LS$-coupling and intermediate coupling  schemes for all three compounds.  This allows us to compare to existing parameters where $LS$-coupling was used, and to investigate the significance of intermediate coupling for all three  materials.  In order to speed up the calculation, the complete basis of the rare earth ions were truncated to the lowest 110 states in \tto{}, 96 states in \dto{}, and 97 states in \hto{}, which still span an energy range of several eV above the CEF ground state.  

\begin{figure}
\includegraphics[scale=1]{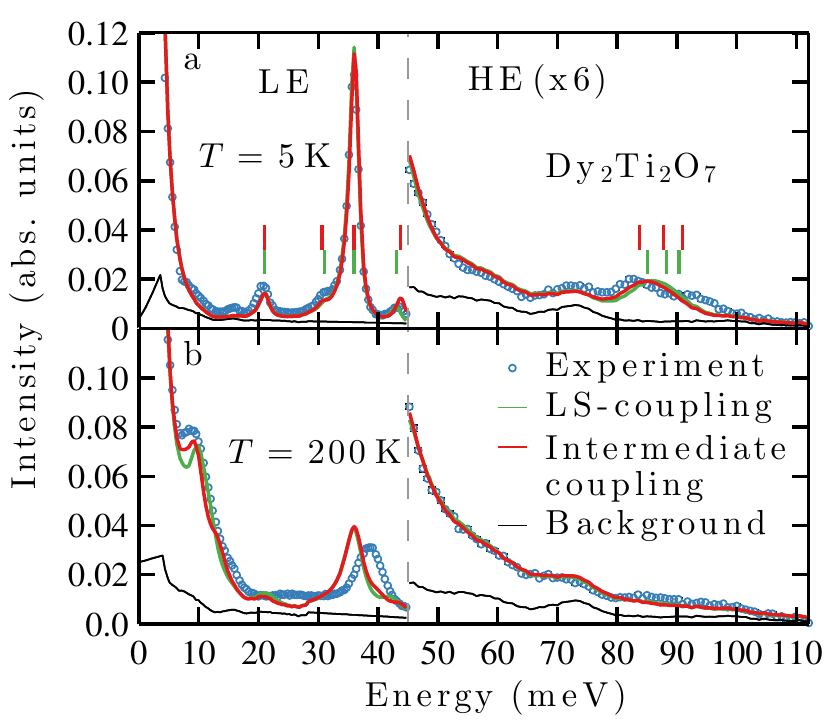}
\caption{Comparison of the observed and calculated inelastic neutron spectra of polycrystalline \dto{} at 5\,K and 200\,K. The spectrum is concatenated from measurements in low energy (LE, $E_i=55$\,meV) and high energy (HE, $E_i=150$\,meV, multiplied by a factor of 6) settings. The theoretical spectra are calculated from the fitted CEF Hamiltonian  and convoluted with the instrumental resolution. At $T=200$\,K, the excitation peaks are broadened with a normalized Lorentzian of width $2$\,meV to account for thermal broadening observed in the experiment.}
\label{fig:dto_cef}
\end{figure}

\begin{figure}
\includegraphics[scale=1]{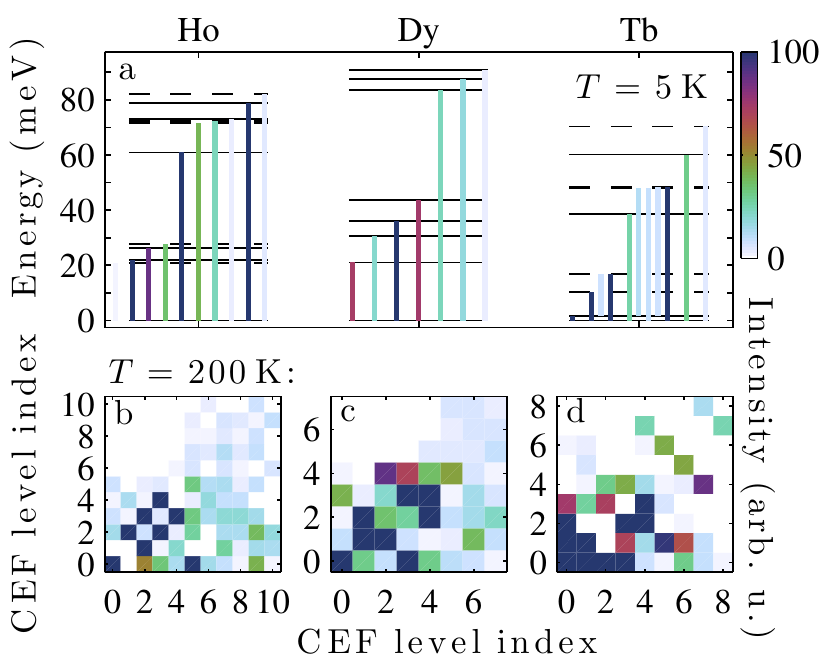}
\caption{Summary of CEF excitations of the \rto{} series. Panel a: The neutron intensity of dipolar transitions between the CEF states at $T=5$\,K are presented in a color code. CEF doublet (singlet) states are illustrated by horizontal solid (broken) black lines. Panels b-d: Color-coded neutron intensity of transitions between thermally excited CEF states at $T=200$\,K. Initial (final) CEF levels are located on the horizontal (vertical) axes.}
\label{fig:cef_summary}
\end{figure}

In a first step, we refined the CEF Hamiltonian of \tto{}, using the CEF parameters from the most recent analysis~\cite{Princep:2015kt} as starting values. Subsequently, the obtained $B_q^k$ parameters were scaled using the point-charge relation of Hutchings~\cite{Hutchings:1964} to serve as starting values for fits of \hto{} and \dto{}. Excellent agreement with the inelastic neutron scattering data was reached in the intermediate coupling scheme with $\chi^2=1.3$ ($R$ = Tb), $\chi^2=1.7$ ($R$ = Dy), and $\chi^2=0.8$ ($R$ = Ho); and good agreement  in the $LS$-coupling scheme with $\chi^2=3.9$ ($R$ = Tb),  $\chi^2=2.2$ ($R$ = Dy), and $\chi^2=1.5$ ($R$ = Ho), where $\chi^2$ represents the standard variance.  The converged $B_q^k$ values are robust against various perturbations of the starting parameters, but sensitively depend upon the experimental observations.  From the best-fit parameters, the energies and relative intensities at $T = 5$ K of the CEF spectra of the three compounds are calculated and summarized in Table~\ref{tab:cf_fitting}. Good agreement is reached with the experimental observations, reflecting the low values of $\chi^2$ mentioned above. Fig.~\ref{fig:dto_cef} shows an example of the close agreement between calculation and experiment for \dto{}. (The calculated CEF transitions were convoluted with the instrumental peak shapes, added to the estimated non-magnetic background and supplemented with an elastic line.  For data at 200 K, a Lorentizian broadening of width 2 meV was included to simulate the thermal broadening of the excitations.) The resulting best-fit crystal field parameters for each compound are summarized in Table~\ref{tab:cf_parameters} and compared with values from other sources.  The wavefunctions of the ground state doublets of each compound, in each coupling scheme investigated are shown in Table~\ref{tab:wf_parameters}.  The level schemes and matrix elements for transitions amongst excited states are summarized graphically in Fig.~\ref{fig:cef_summary}.

\begin{table} 
  \caption{ \label{tab:cf_parameters} Summary of the final CEF parameters $B_q^k$ (in units of meV) for all three rare earth titanate pyrochlores and comparison with the parameters published in the present literature.}
  \centering
  \ra{1.3}
  \begin{tabular}{ r c c c c c c }\hline\hline
     & $B_0^2$ & $B_0^4$ & $B_3^4$ & $B_0^6$ & $B_3^6$ & $B_6^6$\\  \hline
  \multicolumn{1}{l}{\hto{}\phantom{aaaa}} & & & & & & \\
  Ref. \cite{Rosenkranz:2000te} & 68.2 & 275 & 83.7 & 86.8 & -62.5 & 102 \\
  Ref. \cite{Bertin:2012gu} & 61.2 & 271 & 103 & 91.5 & -74.2 & 82.2\\
  LS-coupling & 70.3 & 280 & 81.5 & 87.4 & -62.2 & 108 \\
  Intermediate coupling & 78.2 & 285 & 121 & 113 & -80.8 & 106 \\ \hline
  \multicolumn{1}{l}{\dto{}\phantom{aaaa}} & & & & & & \\
  Ref. \cite{Bertin:2012gu} & 63.0 & 298 & 109 & 102 & 82.2 & 91.5 \\
  LS-coupling & 54.2 & 282 & 187 & 70.1 & 27.7 & 75.1 \\
  Intermediate coupling & 67.5 & 268 & 153 & 66.7 & 77.2 & 132  \\ \hline
  \multicolumn{1}{l}{\tto{}\phantom{aaaa}} & & & & & & \\ 
   Ref. \cite{Gingras:2000vd} & 53.6 & 318 & 146 & 149 & -143 & 67.6\\
   Ref. \cite{Mirebeau:2007hi} & 60.9 & 291 & 103 & 96.6 & -59.9 & 97.5 \\
   Ref. \cite{Klekovkina:hc} & 56.0 & 329 & 95 & 107 & -77.4 & 109 \\
   Ref. \cite{Bertin:2012gu} & 67.3 & 320 & 119 & 113 & -90.5 & 101 \\
   Ref. \cite{Zhang:2014ef} & 144 & 268 & 162 & 171 & 349 & 799 \\
   Ref. \cite{Princep:2015kt} & 55.3 & 370 & 128 & 114 & -114 & 120 \\ 
   LS-coupling &  55.9   &  310   &  114 & 64.6  & -84.2  & 129 \\
   Intermediate coupling & 53.4 & 365 & 108 & 97.8 & -118 & 131 \\ \hline\hline
  \end{tabular}
\end{table}
      
\begin{table*} 
  \caption{ \label{tab:wf_parameters} Wavefunctions of the ground state doublets of the three materials in different coupling schemes.}
  \begin{minipage}{1.0\textwidth}
\flushleft{\hto{}, $LS$-coupling:} \\ 
 \begin{IEEEeqnarray}{rCl}
  |^5I_8,\pm>& = &0.981|^5I_8,\pm8> \mp 0.154 |^5I_8,\pm5> +0.075 |^5I_8,\pm2> \mp 0.073 |^5I_8,\mp1> +0.054 |^5I_8,\mp4> \mp 0.007 |^5I_8,\mp7> \nonumber
 \end{IEEEeqnarray}
\flushleft{\hto{}, intermediate coupling:} \\
 \begin{IEEEeqnarray}{rCl}
 |^5I_J,\pm> & = & 0.978|^5I_8,\pm8> \mp 0.188 |^5I_8,\pm5> +0.027 |^5I_8,\pm2> \mp 0.072 |^5I_8,\mp1> +0.036 |^5I_8,\mp4> \mp 0.006 |^5I_8,\mp7> \nonumber \\
 & & -0.019 |^5I_7,\pm5> \nonumber
  \end{IEEEeqnarray} 
\flushleft{\dto{}, $LS$-coupling:} \\ 
\begin{IEEEeqnarray}{rCl}
|^6H_{15/2},\pm> &=& 0.991|^6H_{15/2},\pm15/2> \mp 0.127 |^6H_{15/2},\pm9/2> +0.019 |{^6H}_{15/2},\pm3/2>\mp 0.025 |^6H_{15/2},\mp3/2> \nonumber \\
& & +0.005 |^6H_{15/2},\mp9/2> \nonumber
\end{IEEEeqnarray}
\dto{}, intermediate coupling: \\
\begin{IEEEeqnarray}{rCl}
|^6H_{J},\pm>& = & 0.996|^6H_{15/2},\pm15/2> \mp 0.001 |^6H_{15/2},\pm9/2> -0.076 |{^6H}_{15/2},\pm3/2>\pm 0.003 |^6H_{15/2},\mp3/2> \nonumber \\
 & & +0.010 |^6H_{15/2},\mp9/2> \nonumber \\
 & & \mp 0.018 |^6H_{13/2},\pm3/2>\mp 0.030 |^6H_{11/2},\pm 9/2>-0.026|^6H_{9/2},\pm9/2>  \mp 0.016 |^6H_{9/2},\pm3/2>\nonumber \\
 & &  -0.017 |^6H_{7/2},\pm3/2> \nonumber
\end{IEEEeqnarray} 
\flushleft{\tto{}, $LS$-coupling:}\\ 
\begin{IEEEeqnarray}{rCl}
|^7F_6,\pm> & = &0.912|^7F_6,\pm4> \mp 0.119 |^7F_6,\pm1>  +0.176 |^7F_6,\mp2> \pm 0.352 |^7F_6,\mp5> \nonumber
\end{IEEEeqnarray}
\flushleft{\tto{}, intermediate coupling:}\\
\begin{IEEEeqnarray}{rCl}
|^7F_J,\pm> & =&  0.967|^7F_6,\pm4> \mp 0.065 |^7F_6,\pm1>  +0.117 |^7F_6,\mp2> \pm 0.191 |^7F_6,\mp5> \nonumber\\
 & &-0.085 |^7F_4,\pm4> +0.021 |^7F_4,\pm2> -0.025 |^7F_5,\pm5> \mp 0.036 |^7F_5,\pm2>\nonumber
 \end{IEEEeqnarray} 
\end{minipage}
\end{table*}
     
\section{\label{sec:Discussion}Crystal field spectra: Discussion}

Various approaches to the determination of CEF Hamiltonian and wavefunction parameters have been taken in these three compounds.  Our aim here was to determine a consistent set of parameters for the three compounds, as well as to clarify certain issues, namely the role of the double peak at $E\approx15.5$ meV in \tto{}, and the exact level scheme of \dto{}.  

The CEF scheme of \hto{}, as determined by Rosenkranz {\it et al.}~\cite{Rosenkranz:2000te} in $LS$-coupling, has been a foundation stone in the study of spin ices.  \hto{} also provides a firm footing for our study.  As described above, the levels we observe are identical to those of Ref.~[\onlinecite{Rosenkranz:2000te}], and when using $LS$-coupling, as was employed in Ref.~[\onlinecite{Rosenkranz:2000te}], the parameters we obtain are in almost exact agreement (see Table.~\ref{tab:cf_parameters}).    The ground state doublet wavefunction is completely dominated by the $|^5I_8,\pm8>$ states, as previously found, and $LS$-coupling is an excellent approximation - generalization to intermediate coupling improves the fit but makes almost no difference to the important contributions to the ground state wavefunctions.

The exact energy levels of \dto{} have not previously been measured, and so their tabulation in Table~\ref{tab:cf_fitting} is in itself a result.  Previous calculations uniformly predict that the first CEF transition has an energy of $E\approx33$ meV, and the highest CEF levels in \dto{} were predicted to have an energy of $E\approx95$ meV~\cite{Qiu:2002}.  Often lower levels are most accurately deduced from bulk measurements or predictions based on other compounds, and we use inelastic neutron scattering to confirm the upper reaches of the CEF scheme.  In this case, we see that the highest level occurs at $E\approx91$ meV, almost exactly as predicted, but the lowest level is at only $E\approx21$ meV, significantly lower than the predicted values.  The predicted Hamiltonian parameters of Ref.~[\onlinecite{Bertin:2012gu}] are nonetheless quite accurate, and the ground state wavefunction is completely dominated by $|^6H_{15/2},\pm15/2\rangle$ states, as expected.  We see that also in the case of \dto{}, the introduction of intermediate coupling does not significantly change the main contributions to the ground state wavefunctions.  

The CEF scheme of \tto{} has been investigated several times, with recent contributions relating to the role of the double peak at $E\approx15.5$ meV and the importance of intermediate coupling.  Zhang {\it et al.}~\cite{Zhang:2014ef} interpreted the double peak as two separate CEF levels, while Princep {\it et al.}~\cite{Princep:2015kt} interpreted it as a single peak split by an unknown mechanism and used the total intensity and energy position of the entire feature.  We modified the level scheme slightly in comparison to that of Ref.~[\onlinecite{Princep:2015kt}] such that  the CEF3 transition was at $E\approx16.7$ meV (the upper part of the double peak) with intensity given only  by the upper part of the double peak.  We will justify this assignment at length below.  This small change in the level scheme does not drastically change the parameters that we obtain, and in Table~\ref{tab:cf_parameters} we see that those of Ref.~[\onlinecite{Princep:2015kt}] and ours (in intermediate coupling) agree very closely.  As discussed below, we have clarified the origin of the double peak, and suggest that the CEF scheme of \tto{} has essentially converged on this intermediate coupling solution with ground state doublet dominated by $|^7F_6,\pm4\rangle$ components, and energy levels as described and tabulated above.

\begin{figure}
\includegraphics[scale=1]{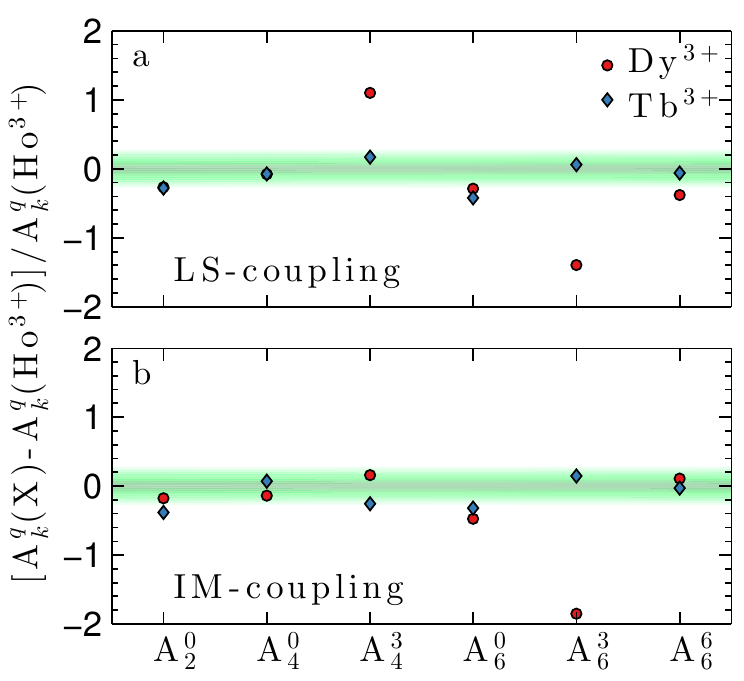}
\caption{Comparison of the $A_l^m$ parameters of the three compounds in $LS$ and intermediate (IM) coupling schemes.  The parameters for \tto{} and \dto{} are displayed relative to those for \hto{}, such that if they were exactly ion independent, they would all appear at zero.  The colored band is a colormap of a normal distribution of width 10 \%.  Note that a parameter of the same magnitude but opposite sign appears at -2 (i.e. $A_6^3$(Dy$^{3+}$ is quite close to this situation).}
\label{fig:cef_scaling}
\end{figure}

Since the three compounds are so closely related, one may expect that the CEF potential should be closely comparable across the series. This may be assessed using the parameters $A_k^q$, which should be ion-independent across the series.  $A_k^q = \lambda_{qk}B_q^k/\langle r^k\rangle$, where $\lambda_{qk}$ is a multiplicative factor specific to each $B_q^k$, and $\langle r^k\rangle$ is the expectation value of the $r^k$ operator, calculated using the Dirac-Fock method (see $\bar{f}$ values in Table V of Ref. ~[\onlinecite{freeman}]).  In Fig.~\ref{fig:cef_scaling} we show $A_k^q$ for \tto{} and \dto{}, normalized to $A_k^q$ for \hto{} each compound.  There is some scatter, but generally the parameters are consistent with this hypothesis, particularly in intermediate coupling.  Noticeable departures occur for $A_4^3$(Dy$^{3+}$) in $LS$-coupling (it is the largest parameter, possibly making the largest relative uncertainties), and $A_6^3$(Dy$^{3+}$) in both coupling schemes, which is of comparable magnitude but opposite sign to $A_6^3$(Tb$^{3+}$, Ho$^{3+}$).  It is explained in Ref.~[\onlinecite{Bertin:2012gu}] that the eigenvalues and eigenvectors of the rare earth pyrochlore crystal field Hamiltonian are not sensitive to the interchange of the sign of $B_3^4$ or $B_3^6$.  This overall consistency suggests that our parameters are reasonable.  A similar point was made in Ref.~[\onlinecite{Princep:2015kt}], where it was pointed out that the parameters for \tto{} obtained there are similar to those for \hto{} and Pr$_2$Sn$_2$O$_7$ after taking account of the difference in radial moments of the $4f$ orbitals, implying similar crystal field potentials amongst rare earth pyrochlore oxides.

We have presented both $LS$-coupling and intermediate coupling parameters, in order to compare to works using either scheme.  Although intermediate coupling does improve the $\chi^2$ of all the fits of the crystal field parameters, and also seems to improve the comparison of $A_k^q$ just mentioned, we do not find it to be equally important for all three compounds.  Intermediate coupling is expected to be most important for \tto{}, due to the largest ratio of ground multiplet splitting and lowest excited multiplet energies (this quantity is $\approx 0.28,0.23,0.13$ for $R=$ Tb, Dy, Ho respectively).  This is born out by the fact that incorporating intermediate coupling has almost no effect on the wave functions of either \dto{} or \hto{}.  We suggest that $LS$-coupling wavefunctions and parameters are completely adequate for \dto{} and \hto{}, but concur with Princep {\it et al.} that intermediate coupling is more appropriate for \tto{}.  The importance of intermediate coupling for \tto{} is visible in the $\chi^2$ values for the parameter fits: for \tto{} the introduction of intermediate coupling produces the largest improvement of $\chi^2$.  This can be understood by reference to the wavefunctions of the excited states, which are tabulated in the Appendix.  In \tto{}, the improvement in the fit is due to the more significant admixture of higher multiplet contributions in excited states on incorporation of intermediate coupling.

The insensitivity of \dto{} and \hto{} to the use of intermediate coupling is also manifested in the ground doublet magnetic moments, which are tabulated in Table~\ref{tab:moments}.  While they are essentially identical for the two coupling schemes in \dto{} and \hto{}, there is a large difference for \tto{}.  The intermediate coupling value of 5.3 $\mu_\mathrm{B}$ is much more comparable with estimates derived from other CEF analyses~\cite{Gingras:2000vd,Mirebeau:2007hi}, and consistent with magnetic field-induced moments observed in neutron diffraction studies~\cite{Cao:2008hw}.  We also report these values as useful quantities.  The magnetic moment of Dy$^{3+}$ and Ho$^{3+}$ in spin ices is involved in the calculation of the monopole charge, and we advance these values as the most appropriate low temperature magnetic moments to use for this task.  The magnetic moment, or, equivalently, the wavefunctions of the groundstate doublets are also important in detailed corrections of susceptibility data for demagnetization effects, where previously they have been approximated as pure $|m_J=\pm J\rangle$ states~\cite{Bovo:2013fq}.

Recent studies of exchange processes and tunneling in \dto{} and \hto{}~\cite{Rau:2015jj,Tomasello:2015vg} have either used a mixture of experimental parameters or interpolations based on literature values for their CEF Hamiltonians, or as a guide to bounds on those quantities.  The assumptions concerning the nature of the groundstate doublet made in these studies seem to be completely justified~\cite{}.  The strength of tunneling depends on sub-leading coefficients in the wavefunctions, and it was noted in Ref.~[\onlinecite{Rau:2015jj}] that, in the case of \dto{}, tunneling is weak, even in the case of the most generous bound of their studied parameters.  The (even smaller) predicted values of the relevant coefficient from Ref.~[\onlinecite{Bertin:2012gu}] would suggest the temperature scale at which tunneling can be important is so low as to be irrelevant to the physics of spin ice.    In our analysis, the coefficient in question (i.e. for $|^6H_{15/2},\mp15/2\rangle$) is effectively zero (there is no contribution to the wavefunction), even smaller again.  Again using values from Ref.~[\onlinecite{Bertin:2012gu}], the temperature scale for \hto{} was found to be even lower.  However, knowledge of all the crystal field energies is important, since their interaction with the phonons may well provide another relaxation channel.  Now that the phonon~\cite{phonons} and CEF spectra are known, we will discuss such processes in a future work.

Ref.~[\onlinecite{Rau:2015jj}] also suggested that current studies may be insufficiently accurate to determine very small parameters well (e.g. the coefficient of $|^5I_8,\pm7\rangle$ in the case of \hto{}).  It is difficult to quantify the accuracy of CEF parameters and wavefunction coefficients, either those obtained in an individual study by fitting of some particular data set, or more generally amongst parameter sets obtained from different techniques or with different interpretations.  The comparison of our parameters for \hto{} with those of Ref.~[\onlinecite{Rosenkranz:2000te}], as in Table~\ref{tab:cf_parameters}, suggests parameters obtained from neutron scattering experiments with comparable energy coverage and congruent interpretations of the excitations, are accurate within $\approx5-10$ \%.  Tests of the sensitivity of the fit to variation of individual parameters suggest a similar level of accuracy for the parameters within this study.  As can be seen from Table~\ref{tab:cf_parameters}, the case of \tto{} is ostensibly less favorable, but it is important to remember that not all of the tabulated studies actually have the same energy range (or interpretation of the excitations).  If more accurate parameters are required, it would be interesting to devise more specific tests.

\begin{table} 
  \caption{ \label{tab:moments} The ground state double magnetic moments in the two coupling schemes (in units of $\mu_\mathrm{B}$ atom$^{-1}$).}
  \centering
  \ra{1.3}
  \begin{tabular}{ c c c }\hline\hline
     & $LS$ & Intermediate \\  \hline
\tto{} & 3.99 & 5.3  \\
\dto{} & 9.92 & 9.93  \\
\hto{} & 9.77 & 9.76  \\
\hline\hline
  \end{tabular}
\end{table}

\section{\label{sec:boundstate_results}Bound state in \tto{}: Results}

\begin{figure}
\includegraphics[scale=1]{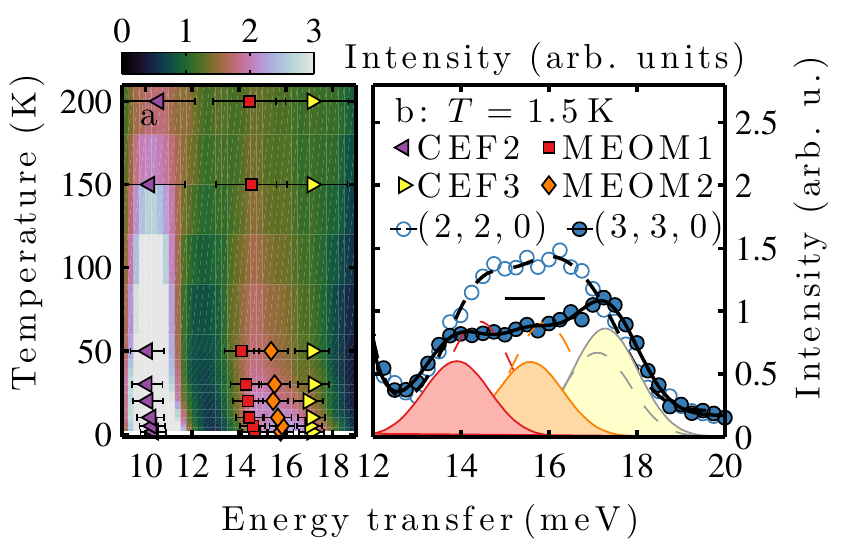}
\caption{Investigations of the bound state in \tto{} by triple axis spectroscopy experiments on the CEF3 transition in a single crystal. Panel a: Colormap representation of the temperature dependence of the inelastic neutron excitation spectrum measured at the Brillouin zone center $\vec{Q}=(2,2,0)$. The extracted excitation centers (widths) are presented by the points (errorbars). Panel b: High resolution scan of the broad envelope including the CEF3 transition at 1.5\,K for $\vec{Q}=(2,2,0)$ and $\vec{Q}=(3,3,0)$, where the latter is the Brillouin zone boundary. We find that the envelope consists of three modes, the CEF3 transition at highest energy, and two magnetoelastic optic modes (MEOM1 and MEOM2), which derive from a bound state between the CEF3 transition and the dispersionless optical phonon at $E\approx15$\,meV.}
\label{fig:tto_meom_raw}
\end{figure}

To obtain our CEF parameters for \tto{}, we assigned only the second peak of the double peak feature as a CEF level.  This assignment was based on the results of separate experiments on a single crystal which we present here.  Fig.~\ref{fig:tto_meom_raw} summarizes our thermal neutron TAS measurements of the energy spectrum in the range $8<E<20$ meV, including the CEF2 and CEF3 excitations. We find that at low temperature, the broad envelope in the excitation spectrum around CEF3 (which appears as a double peak in the powder measurements) actually contains three modes at both the BZ center and boundary ($(2,2,0)$ and $(3,3,0)$ respectively).  A slight spreading of the modes between the BZ center and boundary is visible, but we did not measure their dispersion in detail.  As the temperature increases,  the excitation at the highest energy transfer remains well separated from the other two modes, which eventually appear to merge into one mode for temperatures above $T=100$ K.  We assign the highest mode as CEF3, and propose that the other two excitations originate from a bound state between the CEF3 excitation and the nearly dispersionless transverse optic phonon (TOP) mentioned above, forming two magnetoelastic optic modes (MEOM1 and MEOM2). (The almost exact absence of dispersion of the $E\approx 15$ meV TOP in rare earth titanates can also be seen in our recent investigation of the phonons~\cite{phonons}.)  The observed position of CEF3 on EIGER is $17.1\pm0.2$ meV at $(2,2,0)$, measured with the high resolution PG004 setting , while in the time of flight data it is $16.7\pm0.4$ meV.    In the following, we use $E_{\mathrm{CEF3}}=17.0$ meV, which falls within the error bar of both measurements, and does not qualitatively change any part of the arguments.

\begin{figure}
\includegraphics[scale=1]{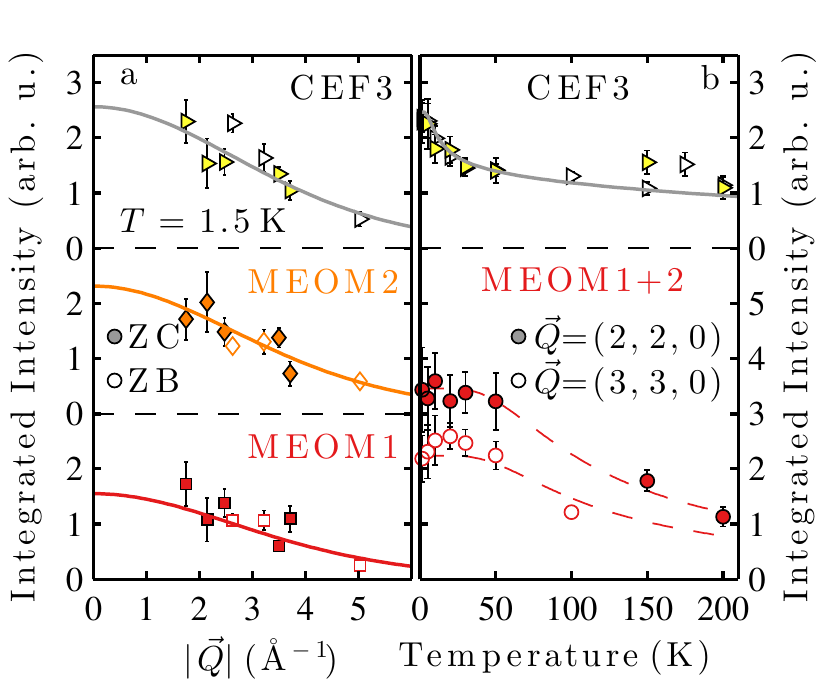}
\caption{Wave vector and temperature dependencies of the components of the bound state in \tto{}. Panel a: Each of the three excitations follows the dipole form factor (solid lines) of the Tb$^{3+}$ ion. Panel b: The temperature dependencies disentangle the nature of the three modes: The excitation at highest energy is evidently the CEF3 transition, as its intensity follows precisely the calculated temperature dependence (solid line). The sum of the integrated intensity of the two MEOMs is related to the population of the CEF3 state in a coupled 4-level system including the ground state and the bound state (dashed lines).}
\label{fig:tto_meom_qt_dep}
\end{figure}

Fig.~\ref{fig:tto_meom_qt_dep} summarizes the extracted temperature and wave vector dependencies of the three components of the broad envelope. All three modes follow the dipole form factor of the Tb$^{3+}$ ion (Fig.~\ref{fig:tto_meom_qt_dep}a), establishing their magnetic character, as shown in Fig.~\ref{fig:tto_meom_qt_dep}a (the zone center (ZC) wave vectors are $(2,2,0)$, $(4,4,0)$, $(0,0,4)$, $(0,0,6)$, $(2,2,2)$, and $(3,3,3)$; the zone boundary (ZB) wavevectors are $(3,3,0)$ and $(1,1,8)$).  Qualitatively different temperature dependencies of the intensities of the first two and the third mode indicate differences in their natures (Fig.~\ref{fig:tto_meom_qt_dep}b). In particular, we found that the intensity of the third mode follows precisely the temperature dependence expected for the CEF3 transition in the CEF spectrum of \tto{} (as established by calculating the intensities as a function of temperature in Spectre and scaling the resulting curve to the data by a single scale factor). The temperature dependence of the other two excitations, in contrast, is best described by a level system that includes only the CEF ground state and the excitations of the envelope, as if decoupled from the rest of the CEF spectrum (i.e. $I\propto n_{E=0}-n_{E=17.0}$, where $n_E=\exp(-E/k_BT)/(1+\exp(-15.1/k_BT)+\exp(-17.0/k_BT))$, with the two MEOMs are represented by a single mode centered at $E=15.1$ meV).  We therefore assigned the third mode of the envelope (which is the second peak of the double peak in the powder spectrum) as the CEF3 transition of the Tb$^{3+}$ ion in \tto{}.  

\section{Bound state in \tto{}: Discussion}

We have shown that the singlet state associated with the CEF3 transition can be assigned to the highest energy mode in the envelope centered at $E\approx15.5$ meV, and propose that the other two modes are magnetoelastic optical modes (MEOMs) - a hybrid excitation carrying both spin and lattice fluctuations.  Let us first systematically exclude other possible origins of these two modes. 

Firstly, these excitations cannot be either pure phonon or CEF transitions, as concluded from their incompatible wave vector and temperature dependencies respectively. Similarly, their excitation energies and temperature dependence (i.e. observable at low temperature) show they cannot originate from transitions between excited  (CEF) states. Moreover, it can be excluded that the two MEOMs appear due to a dynamical Jahn-Teller effect (DJTE)~\cite{Gehring:1975uk}, since both the CEF2 and CEF3 transitions are to singlet states which cannot be split.  We also exclude that the two modes derive from simple magneto-vibrational scattering, which originates from the movement of the electronic cloud of the magnetic ions following the nuclei when they oscillate around their equilibrium position in a phonon~\cite{roessli_chapter}. The neutron cross section of magneto-vibrational scattering follows a combination of the magnetic form factor and the coherent inelastic nuclear cross section $(\vec{Q}\cdot\vec{e})/\omega$, resulting in a selection rule, which is identical to the one for lattice vibrations~\cite{squires}. Our experimental observations therefore rule out magneto-vibrational scattering, because the transverse polarization of the TOP would suppress phonon intensity at longitudinal positions such as $\vec{Q}=(3, 3, 0)$. (Density functional theory (DFT) calculations show that the pure TOP has zero phonon intensity at the longitudinal BZ boundary $\vec{Q}$-point $(3, 3, 0)$, but relatively strong signal at its transverse complement $\vec{Q} = (1, 1, 8)$~\cite{phonons}.)  Additionally, no sign of similar features were observed in the isostructural \hto{} and \dto{}, with close to identical phonon band structures.

Instead, we propose that these MEOMs are microscopic consequences of the spin-lattice coupling involving the quadrupolar oscillator strength of the CEF3 transition and transverse optical phonons. Upon entering the low temperature spin liquid regime below $T=30$ K, \tto{} displays a wealth of magnetoelastic effects, whose microscopic origin is thought to lie in the mixing of CEF and phonon states, as in the formation of a magnetoelastic mode (MEM) by hybridization of the CEF1 doublet and transverse acoustic phonon (TAP)~\cite{Fennell:2014gf}, for example.  

While the MEM follows the dispersion of the TAP but exists only above CEF1, the TOP lies below CEF3, with which it couples, forming two magnetic modes, one of which has an intermediate energy. In agreement with the experimental observation that the intensity of the MEM is related to the population of the CEF1 doublet~\cite{Fennell:2014gf}, we find that the temperature dependence of the MEOM intensities relates to the thermal population of the CEF3 singlet.  

The MEOM proposal is strongly supported by analogy to the microscopically understood bound state physics in CeAl$_2$~\cite{Thalmeier:1982wc} and CeCuAl$_3$~\cite{Adroja:2012cm}. In CeAl$_2$, the result of the coupling between a phonon and a CEF quartet is the splitting of the quartet into two doublets with mixed vibrational and magnetic character, and a renormalized vibrational mode derived from the original phonon. In the neutron cross section, two peaks due to dipole excitations to the two doublets were observed, while in unpolarized Raman scattering the vibrational mode of the coupled system and the lower doublet were detected due to its large vibrational admixture~\cite{Guntherodt:1983io}. The core of the coupling mechanism leading to the predicted bound state lies in large matrix elements for quadrupolar transitions between the ground state and the CEF quartet with identical symmetry and excitation energy as the lattice vibrational modes~\cite{Thalmeier:1982wc}. Similarly in \tto{}, the TOP has compatible energy and symmetry ($E$) to interact with the quadrupolar active CEF3 transition. 

Using the CEF wave functions obtained above, we find large matrix elements of the quadrupolar operators $Q_{xz} = J_xJ_z + J_zJ_x$ and $Q_{yz} = J_yJ_z +J_zJ_y$, which have $E$-symmetry in the $D_{3d}$ point group, between the CEF ground state and the CEF3 state.  The coupling is allowed because $E_g\otimes E_g \otimes A_{jg}\in A_{1g}$ $(j=1,2)$.  In contrast to CeAl$_2$, however, it is not the CEF state that splits in the coupled system, but the TOP, which is doubly degenerate at the BZ center (DFT calculations show that at the BZ center, the double degeneracy of the TOP results from a TO-LO splitting~\cite{phonons}).  At high temperatures, when the states contributing to the broad envelope are thermally populated, the TOP and the CEF state are essentially decoupled and we observe only two modes. The weak MEOM peak persisting up to $T = 200$ K at $\vec{Q} = (2, 2, 0)$ suggests that even higher temperature is needed to fully decouple the states. Upon lowering the temperature, the coupling develops and the TOP splits into two MEOMs with dominant magnetic character. 

From measurements with unpolarized neutrons it is difficult to disentangle a vibrational contribution to the MEOMs. Even at $\vec{Q} = (1, 1, 8)$, where the calculation predicts a sizable phonon cross section for the original TOP, the relevant modes of the envelope appear to have magnetic character, as their intensities follow a similar temperature dependence to the MEOM at (2, 2, 0) and (3, 3, 0) (not shown), and lie on the magnetic form factor. Interestingly, the Raman spectrum measured with perpendicular polarization and therefore sensitive to magnetic fluctuations at the BZ center, shows a similarly broad envelope at the same energies~\cite{Lummen:2008fh}. The observed Raman spectrum, however, differs from the inelastic neutron spectrum in that the intensity of MEOM2 is much reduced compared to the intensity of MEOM1. This result does not conflict with the interpretation of a bound state, but it suggests that MEOM2 develops the most pronounced degree of admixture between magnetic and vibrational fluctuations. 

Having now experimentally characterized the magnetoelastic coupling involving the first~\cite{Fennell:2014gf,in14_in22} and third excited CEF states in \tto{}, the question of whether other CEF states, especially CEF2, can couple similarly to vibrational modes naturally arises. Indeed,  phonon calculations~\cite{phonons} predict optic lattice vibrations with $E$ symmetry at $E\approx10$ meV at the BZ center, dominated by Tb$^{3+}$ ions. Although the same quadrupolar operators relevant for the CEF3 transition have large matrix elements for the CEF2 transition, no magnetoelastic effect is experimentally observed. In contrast to the lattice vibrations involved in the MEM and MEOM (transverse acoustic and transverse optic phonons, respectively), the phonon modes degenerate with the CEF2 singlet have no purely transverse character, which appears to be a necessary ingredient for the magnetoelastic coupling.  We note that a similar effect has been observed in Raman scattering and neutron scattering measurements of Li$R$F$_4$ systems with $R$ = Tb~\cite{Dorfler:1985tc} and very recently with $R$ = Yb~\cite{Babkevich:2015wj}.  Given the complicated effects of applied magnetic field on the excitation spectra of \tto{}~\cite{Rule:2006fr}, we finally note that investigation of the bound state under applied magnetic field promises to be interesting.

\section{\label{sec:Conclusion}Conclusion}

We have measured the crystal field spectra of the three rare earth titanates \rto{}, with $R$ = Tb, Dy, Ho.  We have found a consistent set of parameters for the three compounds, allowing us to accurately parameterize the crystal field wavefunctions of each.  These parameters agree well with other values in the literature where they can be compared.  Furthermore, we have clarified the assignment of optical phonon and crystal field excitations in \tto{} which has been recently debated.  In so doing, we have discovered an example of a bound state between a transverse optical phonon and crystal field state.  Such a bound state has not previously been reported in a rare earth oxide material such as \tto{}, and adds to the catalog of unusual magnetoelastic excitations in this compound.

\acknowledgements{

We acknowledge the important contributions of  M N\'u\~nez Valdez, N A Spaldin and M Kenzelmann to related works, and R Kajimoto for advice on the operation of 4SEASONS.  TF thanks A T Boothroyd and B D Gaulin for discussions, and A L Fennell for highlighting the possibility of performing neutron scattering experiments on natural abundance dysprosium.  Neutron scattering experiments were carried out at the continuous spallation neutron source SINQ at the Paul Scherrer Institut at Villigen PSI in Switzerland; ISIS at the Rutherford Appleton Laboratory, UK; and the MLF of J-PARC, Japan (proposal 2014B0031).  Work at PSI was partly funded by the SNSF (Schweizerischer Nationalfonds zur F\"orderung der Wissenschaftlichen Forschung) (grant 200021\_140862 and 200020\_162626).}

\appendix

\section{Wavefunctions for excited crystal field states}\label{appendix:wavefunctions_for_tto}

In this appendix we present tabulations of the wavefunctions of all the excited crystal field states in the three compounds, in both $LS$-coupling and intermediate coupling schemes.

\begin{table*} 
\caption{Tabulated wave functions of the crystal field states in \hto{} obtained in $LS$-coupling. The crystal field energies are tabulated horizontally, the $m_J$-values of the ground state multiplet vertically. Only coefficients of the wave functions $>10^{-3}$ are shown. For the sake of representation the wave functions of doublet excitations are gathered into one column, of which the values without (in) parentheses correspond to the first (second) member of the doublet.}
 \centering\ra{1.3}\setlength{\tabcolsep}{6pt} 
 \begin{tabular}{r c c c c c c c c c c c} \hline\hline 
$m_J$ & 0.0 & 20.1 & 21.9 & 26.3 & 28.2 & 60.6 & 70.9 & 71.1 & 73.2 & 78.8 & 82.7  \\ \hline 
-8 & 0.981 & & -0.065 & -0.133 & & 0.021 & & & -0.116 & 0.042 & \\ 
-7 & (-0.007) & & (0.042) & (0.157) & & (0.863) & & & (-0.107) & (-0.466) & \\ 
-6 & & -0.065 & & & 0.297 & & 0.323 & -0.704 & & & 0.555 \\ 
-5 & 0.154 & & -0.205 & 0.470 & & -0.259 & & & 0.664 & -0.452 & \\ 
-4 & (0.054) & & (0.643) & (-0.283) & & (0.319) & & & (0.446) & (0.450) & \\ 
-3 & & 0.704 & & & -0.290 & & 0.613 & -0.065 & & & -0.201 \\ 
-2 & 0.075 & & 0.729 & -0.210 & & 0.242 & & & 0.573 & 0.180 & \\ 
-1 & (-0.073) & & (-0.083) & (0.783) & & (0.166) & & & (-0.083) & (0.584) & \\ 
0 & & & & & 0.810 & & 0.202 & & & & -0.550 \\ 
1 & 0.073 & & 0.083 & 0.783 & & 0.166 & & & -0.083 & 0.584 & \\ 
2 & (0.075) & & (0.729) & (0.210) & & (-0.242) & & & (-0.573) & (-0.180) & \\ 
3 & & 0.704 & & & 0.290 & & -0.613 & -0.066 & & & 0.201 \\ 
4 & 0.054 & & 0.643 & 0.283 & & -0.319 & & & -0.446 & -0.450 & \\ 
5 & (-0.154) & & (0.205) & (0.470) & & (-0.259) & & & (0.664) & (-0.452) & \\ 
6 & & 0.065 & & & 0.297 & & 0.322 & 0.704 & & & 0.555 \\ 
7 & 0.007 & & -0.042 & 0.157 & & 0.863 & & & -0.107 & -0.466 & \\ 
8 & (0.981) & & (-0.065) & (0.133) & & (-0.021) & & & (0.116) & (-0.042) & \\ 
 \hline\hline 
 \end{tabular} 
\end{table*}

\begin{table*} 
\caption{Tabulated wave functions of the crystal field states in \hto{} obtained in intermediate coupling. The crystal field energies are tabulated horizontally, the $m_J$-values of the first three multiplets vertically, grouped by multiplet (the multiplets are $^5I_{8}$, $^5I_{7}$, and $^5I_{6}$). Only coefficients of the wave functions $>10^{-3}$ are shown. For the sake of representation the wave functions of doublet excitations are gathered into one column, of which the values without (in) parentheses correspond to the first (second) member of the doublet.}
\centering\ra{1.3}\setlength{\tabcolsep}{6pt} 
 \begin{tabular}{r c c c c c c c c c c c} \hline\hline 
$m_J$ & 0.0 & 20.8 & 21.9 & 26.3 & 27.9 & 61.0 & 71.6 & 72.4 & 73.0 & 78.9 & 82.2  \\ \hline 
-8 & 0.978 & & (-0.019) & -0.149 & & (0.026) & & & (-0.136) & 0.031 & \\ 
-7 & (-0.006) & & 0.031 & (0.115) & & 0.853 & & & -0.032 & (-0.502) & \\ 
-6 & & -0.090 & & & 0.272 & & 0.700 & -0.371 & & & -0.536 \\ 
-5 & 0.188 & & (-0.200) & 0.493 & & (-0.234) & & & (0.723) & -0.318 & \\ 
-4 & (0.036) & & 0.688 & (-0.291) & & 0.310 & & & 0.330 & (0.482) & \\ 
-3 & & 0.701 & & & -0.374 & & 0.090 & -0.565 & & & 0.201 \\ 
-2 & 0.027 & & (0.669) & -0.355 & & (0.216) & & & (0.579) & 0.206 & \\ 
-1 & (-0.072) & & -0.191 & (0.714) & & 0.262 & & & -0.101 & (0.606) & \\ 
0 & & & & & 0.756 & & & -0.290 & & & 0.585 \\ 
1 & 0.072 & & (0.191) & 0.714 & & (0.262) & & & (-0.101) & 0.606 & \\ 
2 & (0.027) & & 0.669 & (0.355) & & -0.216 & & & -0.579 & (-0.206) & \\ 
3 & & 0.701 & & & 0.374 & & 0.090 & 0.565 & & & -0.201 \\ 
4 & 0.036 & & (0.688) & 0.291 & & (-0.310) & & & (-0.330) & -0.482 & \\ 
5 & (-0.188) & & 0.200 & (0.493) & & -0.234 & & & 0.723 & (-0.318) & \\ 
6 & & 0.090 & & & 0.272 & & -0.700 & -0.371 & & & -0.536 \\ 
7 & 0.006 & & (-0.031) & 0.115 & & (0.853) & & & (-0.032) & -0.502 & \\ 
8 & (0.978) & & -0.019 & (0.149) & & -0.026 & & & 0.136 & (-0.031) & \\ 
 \hline
-7 & & & 0.013 & (0.005) & & 0.065 & & & 0.004 & (-0.022) & \\ 
-6 & & 0.010 & & & 0.002 & & 0.002 & -0.013 & & & 0.009 \\ 
-5 & -0.019 & & (0.010) & 0.001 & & (0.011) & & & (-0.017) & 0.020 & \\ 
-4 & (0.001) & & 0.005 & (0.006) & & -0.011 & & & -0.020 & (-0.026) & \\ 
-3 & & 0.016 & & & & & 0.019 & 0.018 & & & -0.034 \\ 
-2 & -0.006 & & (0.018) & 0.002 & & (-0.022) & & & (0.017) & -0.040 & \\ 
-1 & (-0.006) & & 0.002 & (0.009) & & -0.014 & & & 0.040 & (0.017) & \\ 
0 & & 0.002  & & & & & -0.041 & & & & \\ 
1 & -0.006 & & (0.002) & -0.009 & & (0.014) & & & (-0.040) & -0.017 & \\ 
2 & (0.006) & & -0.018 & (0.002) & & -0.022 & & & 0.017 & (-0.040) & \\ 
3 & & -0.016 & & & & & -0.019 & 0.018 & & & -0.034 \\ 
4 & -0.001 & & (-0.005) & 0.006 & & (-0.011) & & & (-0.020) & -0.026 & \\ 
5 & (-0.019) & & 0.010 & (-0.001) & & -0.011 & & & 0.017 & (-0.020) & \\ 
6 & & 0.010 & & & -0.002 & & 0.002 & 0.013 & & & -0.009 \\ 
7 & & & (0.013) & -0.005 & & (-0.065) & & & (-0.004) & 0.022 & \\ 
\hline 
 -6 & & 0.003 & & & 0.005 & & 0.018 & -0.014 & & & -0.011 \\ 
-5 & 0.007 & & (0.004) & 0.003 & & (0.001) & & & (0.010) & 0.001 & \\ 
-4 & & & & (0.006) & & -0.008 & & & -0.005 & (0.004) & \\
-3 & & 0.002 & & & 0.003 & & -0.007 & 0.014 & & & 0.001 \\ 
-2 & 0.006 & & (0.005) & -0.003 & & (-0.006) & & & (-0.009) & -0.009 & \\ 
-1 & (-0.003) & & -0.001 & (0.005) & & 0.010 & & & 0.017 & (-0.005) & \\ 
0 & & & & & 0.004 & & & -0.023 & & & -0.008 \\ 
1 & 0.003 & & (0.001) & 0.005 & & (0.010) & & & (0.017) & -0.005 & \\ 
2 & (0.006) & & 0.005 & (0.003) & & 0.006 & & & 0.009 & (0.009) & \\ 
3 & & 0.002 & & & -0.003 & & -0.007 & -0.014 & & & -0.001 \\ 
4 & & & & -0.006 & & (0.008) & & & (0.005) & -0.004 & \\ 
5 & (-0.007) & & -0.004 & (0.003) & & 0.001 & & & 0.010 & (0.001) & \\ 
6 & & -0.003 & & & 0.005 & & -0.018 & -0.014 & & & -0.011 \\ 
 \hline\hline 
 \end{tabular} 
\end{table*}

\begin{table*} 
\caption{Tabulated wave functions of the crystal field states in \dto{} obtained in $LS$-coupling. The crystal field energies are tabulated horizontally, the $m_J$-values of the ground state multiplet vertically. Only coefficients of the wave functions $>10^{-3}$ are shown. For the sake of representation the wave functions of doublet excitations are gathered into one column, of which the values without (in) parentheses correspond to the first (second) member of the doublet.}
 \centering\ra{1.3}\setlength{\tabcolsep}{6pt} 
 \begin{tabular}{r c c c c c c c c c c c} \hline\hline 
$m_J$ & 0.0 & 0.0 & 21.1 & 31.1 & 31.1 & 36.0 & 43.1 & 85.1 & 88.2 & 90.3 & 90.3   \\ \hline 
-15/2 & 0.991 & & & -0.077 & -0.017 & & & & & 0.104 & 0.003 \\ 
-13/2 & & & 0.149 & & & 0.901 & (-0.180) & 0.330 & (-0.156) & & \\ 
-11/2 & & & (0.046) & & & (0.089) & -0.618 & (-0.393) & 0.674 & & \\ 
-9/2 & -0.127 & 0.005 & & -0.413 & 0.006 & & & & & 0.902 & -0.001 \\ 
-7/2 & & & -0.387 & & & -0.279 & (0.055) & 0.763 & (-0.433) & & \\ 
-5/2 & & & (-0.340) & & & (-0.056) & 0.716 & (-0.188) & 0.578 & & \\ 
-3/2 & 0.019 & -0.025 & & 0.855 & -0.305 & & & & & 0.396 & 0.137 \\ 
-1/2 & & & 0.843 & & & -0.315 & (-0.264) & 0.346 & (0.025) & & \\ 
1/2 & & & (0.843) & & & (-0.315) & 0.264 & (-0.346) & 0.025 & & \\ 
3/2 & 0.025 & 0.019 & & 0.305 & 0.855 & & & & & 0.137 & -0.396 \\ 
5/2 & & & 0.340 & & & 0.056 & (0.716) & -0.187 & (-0.578) & & \\ 
7/2 & & & (0.387) & & & (0.279) & 0.055 & (0.762) & 0.433 & & \\ 
9/2 & 0.005 & 0.127 & & 0.006 & 0.413 & & & & & 0.001 & 0.902 \\ 
11/2 & & & 0.046 & & & 0.089 & (0.618) & 0.393 & (0.674) & & \\ 
13/2 & & & (0.149) & & & (0.901) & 0.180 & (-0.330) & -0.156 & & \\ 
15/2 & & 0.991 & & 0.017 & -0.077 & & & & & 0.003 & -0.104 \\ 
 \hline\hline 
 \end{tabular} 
\end{table*}

\begin{table*} 
\caption{Tabulated wave functions of the crystal field states in \dto{} obtained in intermediate coupling. The crystal field energies are tabulated horizontally, the $m_J$-values of the first three multiplets vertically, grouped by multiplet (the multiplets are $^6H_{15/2}$, $^6H_{13/2}$, and $^6H_{11/2}$). Only coefficients of the wave functions $>10^{-3}$ are shown. For the sake of representation the wave functions of doublet excitations are gathered into one column, of which the values without (in) parentheses correspond to the first (second) member of the doublet.}
 \centering\ra{1.3}\setlength{\tabcolsep}{6pt} 
 \begin{tabular}{r c c c c c c c c c c c} \hline\hline 
$m_J$ & 0.0 & 0.0 & 21.0 & 30.6 & 30.6 & 36.0 & 43.7 & 83.7 & 87.8 & 90.9 & 90.9  \\ \hline 
-15/2 & 0.996 & & & 0.062 & -0.033 & & & & & 0.031 & 0.007 \\ 
-13/2 & & & -0.134 & & & 0.904 & (-0.308) & (0.208) & (0.132) & & \\ 
-11/2 & & & (-0.054) & & & (0.167) & 0.655 & -0.530 & 0.499 & & \\ 
-9/2 & 0.001 & 0.010 & & -0.479 & -0.006 & & & & & 0.874 & -0.001 \\ 
-7/2 & & & -0.372 & & & -0.288 & (0.005) & (0.626) & (0.618) & & \\ 
-5/2 & & & (-0.339) & & & (-0.244) & -0.632 & -0.275 & 0.587 & & \\ 
-3/2 & -0.076 & 0.003 & & 0.772 & -0.400 & & & & & 0.425 & 0.225 \\ 
-1/2 & & & 0.848 & & & -0.068 & (-0.256) & (0.451) & (0.026) & & \\ 
1/2 & & & (0.848) & & & (-0.068) & -0.256 & -0.451 & -0.026 & & \\ 
3/2 & -0.003 & -0.076 & & 0.400 & 0.772 & & & & & 0.225 & -0.425 \\ 
5/2 & & & 0.339 & & & 0.244 & (0.632) & (-0.275) & (0.587) & & \\ 
7/2 & & & (0.372) & & & (0.288) & -0.005 & 0.626 & 0.618 & & \\ 
9/2 & 0.010 & -0.001 & & -0.006 & 0.479 & & & & & 0.001 & 0.874 \\ 
11/2 & & & -0.054 & & & 0.167 & (0.655) & (0.530) & (-0.499) & & \\ 
13/2 & & & (-0.134) & & & (0.904) & -0.308 & -0.208 & -0.132 & & \\ 
15/2 & & 0.996 & & 0.033 & 0.062 & & & & & 0.007 & -0.031 \\ 
 \hline
-13/2 & & & -0.019 & & & 0.062 & (-0.021) & (0.022) & (0.019) & & \\ 
-11/2 & & & (-0.016) & & & 0.009 & 0.039 & -0.025 & 0.033 & & \\ 
-9/2 & 0.005 & 0.002 & & -0.033 & -0.011 & & & & & 0.003 & 0.007 \\ 
-7/2 & & & -0.038 & & & 0.027 & (-0.019) & (-0.018) & (-0.034) & & \\ 
-5/2 & & & -0.038 & & & 0.012 & 0.070 & -0.014 & -0.031 & & \\ 
-3/2 & 0.018 & 0.003 & & -0.063 & -0.030 & & & & & 0.005 & -0.032 \\ 
-1/2 & & & -0.025 & & & 0.014 & (-0.039) & (-0.011) & (0.021) & & \\ 
1/2 & & & 0.025 & & & -0.014 & 0.039 & -0.011 & 0.021 & & \\ 
3/2 & 0.003 & -0.018 & & -0.030 & 0.063  & & & & & 0.032 & 0.005 \\ 
5/2 & & & -0.038 & & & 0.012 & (0.070) & (0.014) & (0.031) & & \\ 
7/2 & & & (-0.038) & & & (0.027) & -0.019 & 0.018 & 0.034 & & \\ 
9/2 & -0.002 & 0.005 & & 0.011 & -0.033  & & & & & 0.007 & -0.003 \\ 
11/2 & & & 0.016 & & & -0.009 & (-0.039) & (-0.025) & (0.033) & & \\ 
13/2 & & & (0.019) & & & (-0.062) & 0.021 & 0.022 & 0.019 & & \\ 
\hline
-11/2 & & & (-0.002) & & & (-0.009) & -0.031 & 0.019 & -0.013 & & \\ 
-9/2 & 0.030 & & & 0.025 & -0.008 & & & & & -0.016 & 0.006 \\ 
-7/2 & & & 0.012 & & & 0.013 & (-0.012) & (0.013) & (-0.001) & & \\ 
-5/2 & & & 0.006 & & & -0.008 & -0.017 & -0.020 & 0.008 & & \\ 
-3/2 & -0.009 & -0.001 & & 0.037 & -0.005 & & & & & 0.013 & -0.009 \\ 
-1/2 & & & 0.047 & & & -0.007 & (-0.006) & (-0.001) & (0.007) & & \\ 
1/2 & & & (0.047) & & & (-0.007) & -0.006 & 0.001 & -0.007 & & \\ 
3/2 & 0.001 & -0.009 & & 0.005 & 0.037 & & & & & -0.009 & -0.013 \\ 
5/2 & & & -0.006 & & & 0.008 & (0.017) & (-0.020) & (0.008) & & \\ 
7/2 & & & (-0.012) & & & (-0.013) & 0.012 & 0.013 & -0.001 & & \\ 
9/2 & & -0.030 & & -0.008 & -0.025 & & & & & -0.006 & -0.016 \\ 
11/2 & & & -0.002 & & & -0.009 & (-0.031) & (-0.019) & (0.013) & & \\  
 \hline\hline 
 \end{tabular} 
\end{table*}

\begin{table*} 
\caption{Tabulated wave functions of the crystal field states in \tto{} obtained in $LS$-coupling. The crystal field energies are tabulated horizontally, the $m_J$-values of the ground state multiplet vertically. Only coefficients of the wave functions $>10^{-3}$ are shown. } 
 \centering\ra{1.3}\setlength{\tabcolsep}{6pt} 
 \begin{tabular}{r c c c c c c c c c c c c c} \hline\hline 
$m_J$ & 0.0 & 0.0 & 1.6 & 1.6 & 10.4 & 16.2 & 44.1 & 44.1 & 45.3 & 45.7 & 62.9 & 62.9 & 69.8  \\ \hline 
-6 & & & & & 0.246 & 0.300 & 0.663 & 0.640 & & & & & -0.006 \\ 
-5 & 0.352 & & 0.891 & & & & & & -0.286 & & -0.027 & & \\ 
-4 & & 0.912 & & 0.374 & & & & & & -0.061 & & -0.158 & \\ 
-3 & & & & & 0.663 & 0.631 & -0.246 & -0.297 & & & & & -0.118 \\ 
-2 & 0.176 & & 0.232 & & & & & & 0.951 & & -0.102 & & \\ 
-1 & & 0.119 & & 0.108 & & & & & & -0.100 & & 0.982 & \\ 
0 & & & & & & 0.155 & & -0.063 & & & & & 0.986 \\ 
1 & -0.119 & & 0.109 & & & & & & 0.100 & & 0.982 & & \\ 
2 & & 0.176 & & -0.232 & & & & & & 0.951 & & 0.102 & \\ 
3 & & & & & 0.663 & -0.631 & -0.246 & 0.297 & & & & & 0.118 \\ 
4 & 0.912 & & -0.375 & & & & & & -0.061 & & 0.158 & & \\ 
5 & & -0.352 & & 0.891 & & & & & & 0.286 & & -0.027 & \\ 
6 & & & & & -0.246 & 0.300 & -0.663 & 0.640 & & & & & -0.006 \\ 
 \hline\hline 
 \end{tabular} 
\end{table*}

\begin{table*} 
\caption{Tabulated wave functions of the crystal field states in \tto{} obtained in intermediate coupling. The crystal field energies are tabulated horizontally, the $m_J$-values of the first three multiplets vertically, grouped by multiplet (the multiplets are $^7F_6$, $^7F_5$, and $^7F_4$). Only coefficients of the wave functions $>10^{-3}$ are shown. } 
  \centering
  \ra{1.3}
  \setlength{\tabcolsep}{6pt}
  \begin{tabular}{r c c c c c c c c c c c c c }\hline\hline
  \multicolumn{13}{c}{Energy (meV)} \\
  $m_J$ & 0.0 & 0.0 & 1.5 & 1.5 & 10.2 & 17.0 & 38.7 & 38.7 & 47.9 & 48.4 & 60.2 & 60.2 & 70.4 \\ \hline
-6 & & & & & 0.130 & 0.168 & & & -0.690 & 0.680 & & & -0.056 \\ 
-5 & & 0.191 & 0.955 & & & & & -0.152 & & & & -0.042 & \\ 
-4 & 0.967 & & & 0.204 & & & -0.087 & & & & -0.072 & & \\ 
-3 & & & & & 0.688 & 0.679 & & & 0.120 & -0.163 & & & -0.039 \\ 
-2 & & 0.117 & 0.142 & & & & & 0.963 & & & & -0.032 & \\ 
-1 & 0.065 & & & 0.061 & & & -0.032 & & & & 0.978 & & \\ 
0 & & & & & & 0.076 & & & & 0.061 & & & 0.984 \\ 
1 & & -0.065 & 0.061 & & & & & 0.032 & & & & 0.978 & \\ 
2 & 0.117 & & & -0.142 & & & 0.963 & & & & 0.032 & & \\ 
3 & & & & & 0.688 & -0.679 & & & 0.120 & 0.163 & & & 0.039 \\ 
4 & & 0.967 & -0.204 & & & & & -0.087 & & & & 0.072 & \\ 
5 & -0.191 & & & 0.955 & & & 0.152 & & & & -0.042 & & \\ 
6 & & & & & -0.130 & 0.168 & & & 0.690 & 0.680 & & & -0.056 \\  \hline
-5 & & -0.025 & -0.137 & & & & & 0.071 & & & & 0.023 & \\ 
-4 & 0.001 & & & -0.001 & & & 0.022 & & & & 0.047 & & \\ 
-3 & & & & & 0.089 & 0.055 & & & 0.067 & -0.061 & & & 0.028 \\ 
-2 & & 0.036 & -0.003 & & & & & 0.173 & & & & 0.009 & \\ 
-1 & 0.004 & & & 0.026 & & & 0.008 & & & & 0.138 & & \\ 
0 & & & & & 0.007 & & & & 0.032 & & & & \\ 
1 & & 0.004 & -0.026 & & & & & 0.008 & & & & -0.138 & \\ 
2 & -0.036 & & & -0.003 & & & -0.173 & & & & 0.009 & & \\ 
3 & & & & & -0.089 & 0.055 & & & -0.067 & -0.061 & & & 0.028 \\ 
4 & & -0.001 & -0.001 & & & & & -0.022 & & & & 0.047 & \\ 
5 & -0.025 & & & 0.137 & & & 0.071 & & & & -0.023 & & \\  \hline
-4 & -0.085 & & & -0.018 & & & 0.019 & & & & 0.043 & & \\ 
-3 & & & & & -0.028 & -0.050 & & & -0.037 & 0.049 & & & 0.035 \\ 
-2 & & 0.021 & -0.013 & & & & & 0.004 & & & & 0.015 & \\ 
-1 & -0.014 & & & 0.025 & & & 0.003 & & & & 0.084 & & \\ 
0 & & & & & & 0.009 & & & & 0.045 & & & 0.119 \\ 
1 & & 0.014 & 0.025 & & & & & -0.003 & & & & 0.084 & \\ 
2 & 0.021 & & & 0.013 & & & 0.004 & & & & -0.015 & & \\ 
3 & & & & & -0.028 & 0.050 & & & -0.037 & -0.049 & & & -0.035 \\ 
4 & & -0.085 & 0.018 & & & & & 0.019 & & & & -0.043 & \\ \hline\hline
  \end{tabular}
\end{table*}


%

\end{document}